\documentclass[12pt]{article}
\usepackage{fullpage}
\usepackage{setspace}
\usepackage[pdftex,colorlinks=true]{hyperref}
\usepackage{url}

\begin{document}

\title{\textbf{Finding and Using Electromagnetic Counterparts of Gravitational 
Wave Sources} \\
\vskip3pt
\sc{\Large A white paper for the Astro2010 Decadal Review}}

\author{E. Sterl Phinney \\
Theoretical Astrophysics\\
130-33 California Institute of Technology \\
Pasadena, CA 91125\\
\texttt{esp@tapir.caltech.edu}
\\[2em]
}
\maketitle

\begin{abstract}
The principal goal of this whitepaper is not so much 
to demonstrate that gravitational wave detectors like LIGO and LISA
will help answer many central questions in astronomy and astrophysics,
but to make the case that they can help answer {\bf a far greater range
of questions} {\em if we prepare to make the (sometimes substantial)
effort to identify electromagnetic counterparts to the gravitational 
wave sources.}
\end{abstract}

\clearpage

\section{Extreme Sports in the Universe: past decades}
\label{sec:extremepast}

The most dramatic events in the universe ---the deaths of stars, the
collisions of stellar remnants and giant black holes, the feeding of the
monsters in galactic nuclei ---are believed to produce the most
electromagnetically luminous objects in the universe: supernovae,
gamma-ray bursts, quasars.

The objects involved in these events are worthy of study for three 
reasons: 1) their
intrinsic interest and the great distances to which they can be seen,
2) the impact of the energy they release on the rest of the 
astronomical universe, and 3) (perhaps most
significantly) because they test our theories of matter and
energy in ways we cannot hope to do on earth.  They involve the most extreme
physical conditions in the present universe: the highest densities of 
both matter and radiation, the highest magnetic fields, 
the deepest gravitational fields, the most relativistic bulk motions.

In previous decades, we have studied these events using just the
light emitted from them (plus a couple of dozen marvelous neutrinos from 
supernova 1987A).  This light comes from atoms and electrons in tenuous gas,
generally far removed from the main action. Because this action
is buried deep behind layers of obscuring gas and in deep gravitational 
potential wells, the light we can see has given us many circumstantial
clues, but the deepest questions remain unanswered. 

\begin{itemize}
\item How and why do some stars explode as supernovae?
\item Are the things we call black holes actually precisely the vacuum space-time
solutions of the equations of general relativity?
\item How and when do black holes form in stars and in the centers of galaxies?
\item Do black holes form in other places?
\item How do gas, stars and black holes interact in the nuclei of galaxies and
compact clusters?
\item What are the internal ingredients of neutron stars?
\item What happens when two white dwarfs merge?
\end{itemize}

\section{The next decade: adding information from gravitational waves}
\label{sec:extremefut}

Strong gravitational waves are produced by the rapid motion of 
massive compact bodies: exactly those extreme objects discussed above.
The waves encode the history of those motions: exactly the information
that has proven so difficult to obtain electromagnetically. 
Gravitational waves will provide us with a detailed look deep into the
interiors of the most exotic objects in our Universe.  
Ground-based detectors such as enhanced-LIGO (2009) and advanced LIGO (2014)
will detect high-frequency (\textbf{HF}) gravitational waves ($\sim 10-1000$Hz)
\cite{advligoprop}.
They can detect the merging of binary black holes, and the tidal disruption 
and merger of neutron stars in black hole and neutron star binaries at
$\sim$50Mpc and $\sim$200Mpc respectively.
They may also be able to detect accretion-induced collapse events,
and some types of supernovae and pulsars \cite{melatosgwacc}.
The ESA-NASA space-based mission LISA will detect low-frequency (\textbf{LF}) 
gravitational waves (0.1-10mHz) \cite{prince-whitepaper, lisadocuments}. 
It can detect
merging binary supermassive black holes (to $z\sim30$),  their
captures of intermediate mass black holes (to $z\sim3$), and 
their captures of the compact objects (stellar mass black holes to $z\sim 1$, 
neutron stars and white dwarfs to $z\sim 0.1$ ) in galactic nuclei.  
It will measure the masses, spins and distances of all these objects
to precisions unprecedented in astrophysics.
It will also measure the orbital parameters of 
thousands of ultracompact binary stars in the Milky Way and its satellites.

From the gravitational waveforms we will be able to decode precise
information about the masses, spins, distances, interior properties
and space-time dynamics of the detected sources.  This information will be
of an accuracy and robustness far exceeding what can be obtained by
electromagnetic measurements.  Alone, these will enable precision tests
of general relativity in the strong-field 
limit\cite{schutz-whitepaper, whitcomb-whitepaper}
and the structure and 
dynamics of compact objects, and tell us about the merger rates as a
function of cosmic time for supermassive black holes\cite{madau-whitepaper}, 
intermediate mass 
black holes, and stellar mass black holes, neutron stars 
and white dwarf binaries of all 
types\cite{miller-whitepaper, nelemans-whitepaper}.

\section{Why electromagnetic counterparts to gravitational wave sources
are so important}
\label{sec:emcounter}

However this new information from gravitational waves
can have a far broader impact if it can
be put into the context of our existing electromagnetic view of the universe.

Imagine the frustration of measuring gravitational waves from what
appears to be the formation of a rapidly spinning neutron star,
but missing the electromagnetic counterpart that would determine that
it was from a massive type Ic supernova in a starburst galaxy, not 
an accretion-induced collapse of an accreting white dwarf
in the outskirts of an elliptical galaxy.

And what a lost opportunity if gravitational waves measured the properties
of a merging pair of massive black holes, but we missed the electromagnetic
fireworks that would have enabled us to identify the host high-redshift
galaxy and its clustering environment, and to
diagnose the properties of the black holes' circumbinary gas disk.

Gravitational waves alone will generally determine sky positions only to
$\sim$ degrees \footnote{Tens of degrees for short-lived LF (LISA)
sources and high-frequency sources (LIGO), improving for sources with
$>$ year lifetimes to $\sim 0.1$ degree for strong LISA sources and
arcseconds for strong LIGO sources.}.
In a few cases, bright electromagnetic events such as nearby gamma-ray
bursts or supernovae may trigger LIGO searches\cite{ligotriggered}.
But most commonly, gravitational wave detections by LIGO and LISA
will have to trigger electromagnetic searches,
because many of the predicted
counterparts are faint (V magnitudes 18-27) and can be short-lived and
inconspicuous. Much effort on the part of electromagnetic observations
will be required to identify them in the error boxes that may be as large as
tens of square degrees.  Wide field synoptic survey instruments with
a fast cadence are ideal (e.g. Palomar Transient Factory, Pan-Starrs, 
SkyMapper, LSST in the optical, plus wide-field or all-sky radio, 
X-ray and gamma-ray instruments\cite{bloom-whitepaper}).  
Finding electromagnetic counterparts will require rapid, but
feasible data analysis and dissemination from the gravitational wave
detectors (within minutes to hours for short-lived HF sources, and
days to weeks for merging LF sources).  

Fortunately, finding counterparts of
both HF and LF gravitational wave sources requires quite similar 
electromagnetic instrumentation, of a type envisaged
for the coming decade.  Flexibility in the cadences and modes of operation
will, however, be required.

In the subsections below, we give just a few selected
examples of science questions that
could be answered by combining electromagnetic and gravitational wave
data on the sources listed. The tables briefly summarise additional topics
for which space precludes discussion.
\footnote{In keeping with US legal tradition ``innocent until proven guilty'',
we use the term ``dead stars'' instead of the more common
terms ``compact objects'' or ``white dwarfs, neutron stars, and black holes'',
which presuppose proof beyond a reasonable doubt.}.

\subsection{What causes some pairs of dead stars to merge, and others not to?}
\label{sec:deadmerger}
The fates of interacting binary stars remain poorly understood, despite 
their importance for the nature, formation and evolution of Type I (a,b,c)
supernovae, binary pulsars, X-ray binaries and cosmic nucleosynthesis.
A priori calculations involve accretion disks, magnetic fields, and
the 3-D radiation hydrodynamics of common-envelope evolution. Parametrised
population synthesis has many uncertainties and difficulties explaining
existing population data, which is however subject to severe selection effects.

LISA will provide a complete Galactic
sample orders of magnitude larger than the handful of 
ultracompact binaries now known, 
measuring the individual parameters of
about 20,000 ultracompact white dwarf binaries (orbital
periods less than around 30 minutes) in the Milky Way and its satellites
\cite{vallisneri-lisadata}, 
along with a less well-determined number of neutron star 
and black hole binaries. The frequency $f$ of the gravitational wave signal
is twice the orbital frequency. 
The polarisation of the gravitational wave signal determines the orbit
inclination and orientation on the sky.  The amplitude of the gravitational
signal $h\propto f^{2/3}\mathcal{M}^{5/3}/D$ 
depends on the chirp mass $\mathcal{M}=(M_1M_2)^{3/5}(M_1+M_2)^{-1/5}$
and distance $D$ to the source.  If the components are not interacting by
mass transfer or tides, the orbital frequency evolution $\dot f$ is determined
just by gravitational radiation (and the 
braking index $\ddot f f/\dot f^2=11/3$), and $D$ can be determined
accurately for the $\sim 3000$ ultracompact binaries for which $\dot f$ will
be measurable. But for many of the white dwarf binaries, tides and
mass transfer will affect $\dot f$, even in sign.  For these, $\mathcal M$
can still be determined from $h$ if $D$ is known.  
GAIA will measure parallaxes for
at least $\sim 400$ (more if tidal heating is important) eclipsing binary
white dwarfs, with substantial overlap with the LISA sample, so for
these sources, the combination of EM and GW information will determine
the component radii, masses and orbital dynamics\cite{stroeer-lisagaia}.
Identifying these eclipsing white dwarf pairs in the $\sim 1^\circ$ error
boxes from LISA will require fast (minute cadence) wide field synoptic surveys
to $V=19$, and followup with high-speed spectroscopy of the identified sources.

\begin{tabular}{p{1.3in} | p{1.3in} | p{3.3in}}
Source & GW data & GW+EM data \\
\hline
ultracompact \quad binaries (LISA) &
complete census in Milky Way and satellites, binary periods, component masses, orbit inclination and orientation, in- or out-spiral rates, formation rate, merger rate &
parallaxes (GAIA) needed for some GW masses, eclipses: white dwarf radii,
white dwarf temperatures: tidal heating?, tidal synchronisation?,
accretion rates and types (disk, direct impact, magnetic guiding), winds,
SN Ia, AIC?\\
\hline
\end{tabular}

\subsection{What happens during the mergers of dead stars, and what determines 
the nature of their remnants?}

Merging neutron star binaries (and/or neutron stars tidally disrupted in
merging neutron star-black hole binaries) have been proposed as the
engines of short-hard gamma-ray bursts, $r$-process nucleosynthesis, and
ultra-high energy neutrinos.
Yet these identifications remain hypothetical for lack of evidence.
Gravitational wave detections will settle the matter. 
But because of relativistic beaming, the majority of mergers may
not be detectable as gamma-ray bursts, and such optical and radio afterglows 
(likely less beamed) as have been seen are faint, and will require concerted
effort from wide-field synoptic survey telescope and much followup to
distinguish them from other transient sources.  Yet the payoff will be
tremendous ---for example, the gravitational waveform will determine
the orbit inclination as well as component masses and radii and the
time of merger.  This will provide vastly improved constraints on models of the
electromagnetic emission as a function of angle from the angular momentum
axis.

Similarly, a combination of gravitational wave and electromagnetic
detections could immediately clarify our uncertainty about what 
happens when two
white dwarfs or a white dwarf and a neutron star accrete and merge: 
outcomes range from neutron stars or black holes (accretion-induced collapse,
AIC) to various explosive ejections up to type Ia supernovae. 

\begin{tabular}{p{1.3in} | p{1.3in} | p{3.3in}}
Source & GW data & GW+EM data \\
\hline
NS-BH (LIGO) &
masses, spins, NS radius/structural parameters, orbit inclination and orientation&
Gamma-rays (GRB?), jet and outflow (afterglow) properties as function of GW-determined inclination, neutrinos, nucleosynthesis.\\
\hline
NS-NS (LIGO) &
masses, neutron star radii/structural parameters &
Gamma-rays (GRB?), jet and outflow (afterglow) properties as function of GW-determined inclination, neutrinos, nucleosynthesis.\\
\hline 
WD-WD, WD-NS (LIGO, LISA) &
  final NS or BH mass, radius, spin &
  nucleosynthesis, outflow energetics, explosion vs remnant?, magnetic field of AIC remnant\\
\hline
\end{tabular}

\subsection{What are the fates of stars and stellar remnants 
near the giant black holes in galactic nuclei?}

Observations of the center of our Milky Way and other nearby galaxies
have revealed that short-lived massive stars form surprisingly close to the 
supermassive black holes that lurk there.  Mass segregation implies
that black hole remnants of such stars should be the dominant population
close to the black hole,
and they would provide valuable clues to the star formation history and
stellar dynamics in galactic nuclei.  Detection of these is only possible
through gravitational radiation.   White dwarfs may be tidally disrupted
late in their inspiral, providing both electromagnetic and gravitational
wave signatures and the combination of gravitational wave determined
distance (see section~\ref{sec:deadmerger}) and electromagnetic
redshift would enable precision cosmography \cite{hogan-whitepapera, haiman}.
These sources also will provide maps of the spacetime geometry of 
unprecedented precision (black hole multipole moments with precision as
high as $\sim 10^{-4}$).  While these enable tests of strong-field general
relativity (\cite{schutz-whitepaper}), comparison of the unambiguously
known masses and spins for a population of quiescent and active black holes
will at last provide a firm basis for electromagnetic models of accretion and
the evolution of black holes in galactic nuclei 
(\cite{miller-whitepaper, madau-whitepaper}).

\begin{tabular}{p{1.3in} | p{1.3in} | p{3.3in}}
Source & GW data & GW+EM data \\
\hline
extreme mass ratio inspiral onto massive black holes in galactic nuclei (LISA) &
 mass, spin of massive black hole, mass of compact object, distance, high-precision spacetime, merger rate and dead star mass function in local universe, orbit eccentricity, inclination, orientation &
 redshift, host galaxy ID, properties and dynamics of host nucleus, GW advance notice for EM study of tidal disruption of white dwarfs, brown dwarfs of known M,R.\\
\hline
\end{tabular}

\subsection{How and when do black holes form in the centers of galaxies?}

The field of electromagnetic counterparts of merging supermassive black
holes has burgeoned in the past few years \cite{milosesp,haiman}.
Signatures range from prompt optical, UV and X-ray flares to year-timescale
events in infrared and other bands.
Identification of these transients will enable us to identify the host
galaxy, testing models of galaxy mergers, and also, because of the known
time and amplitude of disk perturbation around a black hole of known mass,
potentially revolutionise the study of accretion disks, as well
as testing modified gravity models by comparing the electromagnetic
and gravitational $z-D_L(z)$ relations \cite{haiman}.

\begin{tabular}{p{1.3in} | p{1.3in} | p{3.3in}}
Source & GW data & GW+EM data \\
\hline
merging massive or intermediate mass black holes (LISA) &
 masses, spins (initial and final), distances, merger rates, spacetime dynamics  &
 host galaxy morphology, luminosity, dynamics, redshift; circumbinary gas disk properties and response to mass loss, kicks, gravitational waves; magnetodynamics\\
\hline
\end{tabular}

\subsection{How and why do some stars explode as supernovae?}

Most supernovae probably are not strong sources of gravitational waves,
but the most interesting ones (e.g. rapidly rotating hypernovae implicated in
long-duration gamma-ray bursts, magnetar progenitors, etc) may be
\cite{ottsnae}.  These gravitational waveforms will likely
only be identified in coincidence with electromagnetic observations,
but would provide a view of the interior dynamics of supernova unobtainable
in any other way.

\begin{tabular}{p{1.3in} | p{1.3in} | p{3.3in}}
Source & GW data & GW+EM data \\
\hline
Supernovae (LIGO) & 
 extreme core (magneto) hydrodynamics and instabilities, rotation, bars &
 relation of core dynamics to explosion type, progenitors, nucleosynthesis, black hole vs neutron star remnant\\
\hline
\end{tabular}

\subsection{What are the internal ingredients of neutron stars?}

Rapidly rotating, magnetised and accreting neutron stars may not be
axisymmetric.  Gravitational waves can provide one of our few ways
to determine the interior structure and motions\cite{melatosgwacc}.

\begin{tabular}{p{1.3in} | p{1.3in} | p{3.3in}}
Source & GW data & GW+EM data \\
\hline
radio and X-ray pulsars, low-mass X-ray binaries (LIGO) & 
 spin, interior structures,
 neutron star mountains and mass multipole moments induced by interior magnetic fields or accretion &
 accretion rate, spin up/down, exterior magnetic field, detection of core-crust differential rotation, thermal properties\\
\hline
\end{tabular}

\begin{spacing}{0.9}

\end{spacing}
\end{document}